\input{aipcheck}

\documentclass[
    ,final            % use final for the camera ready runs
  ]
  {aipproc}
\layoutstyle{6x9}
\newcommand{\dd}{{\rm d}}

\begin{document}

\title{A precise new KLOE measurement of $|F_\pi|^2$ with ISR events
and determination of $\pi\pi$ contribution
to $a_\mu$ for $0.592 < M_{\pi\pi} < 0.975$ GeV}


\author{G. Venanzoni for the KLOE Collaboration\thanks{F.~Ambrosino,
A.~Antonelli,
M.~Antonelli,
F.~Archilli,
C.~Bacci,
P.~Beltrame,
G.~Bencivenni,
S.~Bertolucci,
C.~Bini,
C.~Bloise,
S.~Bocchetta,
F.~Bossi,
P.~Branchini,
P.~Campana,
G.~Capon,
T.~Capussela,
F.~Ceradini,
F.~Cesario,
S.~Chi,
G.~Chiefari,
P.~Ciambrone,
F.~Crucianelli,
E.~De~Lucia,
A.~De~Santis,
P.~De~Simone,
G.~De~Zorzi,
A.~Denig,
A.~Di~Domenico,
C.~Di~Donato,
B.~Di~Micco,
A.~Doria,
M.~Dreucci,
G.~Felici,
A.~Ferrari,
M.~L.~Ferrer,
S.~Fiore,
C.~Forti,
P.~Franzini,
C.~Gatti,
P.~Gauzzi,
S.~Giovannella,
E.~Gorini,
E.~Graziani,
W.~Kluge,
V.~Kulikov,
F.~Lacava,
G.~Lanfranchi,
J.~Lee-Franzini,
D.~Leone,
M.~Martemianov,
M.~Martini,
P.~Massarotti,
W.~Mei,
S.~Meola,
S.~Miscetti,
M.~Moulson,
S.~M\"uller,
F.~Murtas,
M.~Napolitano,
F.~Nguyen,
M.~Palutan,
E.~Pasqualucci,
A.~Passeri,
V.~Patera,
F.~Perfetto,
M.~Primavera,
P.~Santangelo,
G.~Saracino,
B.~Sciascia,
A.~Sciubba,
A.~Sibidanov,
T.~Spadaro,
M.~Testa,
L.~Tortora,
P.~Valente,
G.~Venanzoni,
R.~Versaci,
G.~Xu.
}}{
address={Laboratori Nazionali di Frascati dell'INFN, Via E. Fermi 40, I-00044 Frascati, Italy \,\,\,\,\,\,\,\,\,\,\,\, \,\,\,\,\,\,\,\,\,\,\,\,\,\,\,\, \,\,\,\,\,\,\,\,\,\,\,\,\,\,\,\,\,\, \emph{e-mail}: graziano.venanzoni@lnf.infn.it}}

\begin{abstract}
The KLOE experiment at the DA$\Phi$NE $\phi$-factory has
performed a new precise measurement of the pion form factor
using Initial State Radiation events, with photons emitted
at small polar angle. Results based on an integrated luminosity
of 240 pb$^{-1}$ and extraction of the
$\pi\pi$ contribution to $a_\mu$ in the mass range $0.35< M^2_{\pi\pi}<0.95$ GeV$^2$
are presented. The new value of  $a^{\pi\pi}_\mu$ has smaller (30\%) statistical and systematic error and is consistent with the KLOE published value 
(confirming the current disagreement between the standard model prediction for $a_\mu$ and the measured value).
\end{abstract}

\maketitle

%%%%%%%%%%%%%%%%%%%%%%%%%%%%%%%%%%%%%%%%%%%%
%% MAINMATTER
%%%%%%%%%%%%%%%%%%%%%%%%%%%%%%%%%%%%%%%%%%%%

\section{Introduction}
`The anomalous magnetic moment of the muon
has recently been measured to an accuracy
of 0.54 ppm~\cite{Bennett:2006fi}.
The main source of uncertainty
in the value predicted~\cite{Jegerlehner:2007xe} in the Standard Model
is given by the hadronic contribution, $a_\mu^{hlo}$,
to the lowest order.
This quantity is estimated with a dispersion
integral of the hadronic cross section measurements.

In particular, the pion form factor, $F_\pi$, defined via
$\sigma_{\pi\pi}\equiv\sigma_{e^+ e^-\to\pi^+\pi^-} = \frac{\pi\alpha^2}{3 s}
\beta^3_\pi(s) |F_\pi(s)|^2$,
accounts for $\sim70\%$ of the central
value and for $\sim60\%$ of the uncertainty
in $a_\mu^{hlo}$.

The KLOE experiment already published~\cite{Aloisio:2004bu}
a measurement of $|F_\pi|^2$ with the method described
below, using an integrated luminosity
of 140 pb$^{-1}$, taken in 2001, henceforth
referred to as KLOE05, with a fractional systematic error of $1.3\%$.

\section{Measurement of $\sigma(e^+e^-\to\pi^+\pi^-\gamma)$ at DA$\Phi$NE}
\label{sec:2}
DA$\Phi$NE is an $e^+ e^-$ collider running %from 1999 to 2005
at $\sqrt{s}\simeq M_\phi$,
the $\phi$ meson mass, which has
provided % collecting
an integrated luminosity of about 2.5 fb$^{-1}$
to the KLOE experiment up to year 2006.
In addition, about 250 pb$^{-1}$ of data have been collected
at $\sqrt{s}\simeq 1$ GeV, in 2006.
Present results are based on 240 pb$^{-1}$
of data taken in 2002 (3.1 Million events)~\cite{Ambrosino:2008en}. 
The KLOE detector consists of a drift chamber~\cite{Adinolfi:2002uk} with
excellent momentum resolution ($\sigma_p/p\sim 0.4\%$
for tracks with polar angle larger than $45^\circ$)
%and vertex ($\sigma_{vtx}\sim3$ mm) resolution
and an electromagnetic calorimeter~\cite{Adinolfi:2002zx} with good energy
($\sigma_E/E\sim 5.7\%/\sqrt{E~[\mathrm{GeV}]}$)
and precise time ($\sigma_t\sim 54~\mathrm{ps}/\sqrt{E~[\mathrm{GeV}]}\oplus
100~\mathrm{ps}$) resolution.

At DA$\Phi$NE, we measure the differential spectrum %cross section
of the $\pi^+\pi^-$ invariant mass, $M_{\pi\pi}$, from
Initial State Radiation (ISR) events,
$e^+ e^-\to\pi^+\pi^-\gamma$, and extract
the total cross section $\sigma_{\pi\pi}\equiv\sigma_{e^+ e^-\to\pi^+\pi^-}$
using the following formula~\cite{Binner:1999bt}:
\begin{equation}
s~ \frac{\dd\sigma_{\pi\pi\gamma}}
{\dd M_{\pi\pi}^2} = \sigma_{\pi\pi}
(M_{\pi\pi}^2)~ H(M_{\pi\pi}^2)~,
\label{eq:1}
\end{equation}
where $H$ is the radiator function.
This formula neglects
Final State Radiation (FSR) terms (which are properly taken into account in the analysis).
% -----
\begin{figure}[h!]
%  \begin{center}
%    \subfigure
    \resizebox{10pc}{!}{\includegraphics{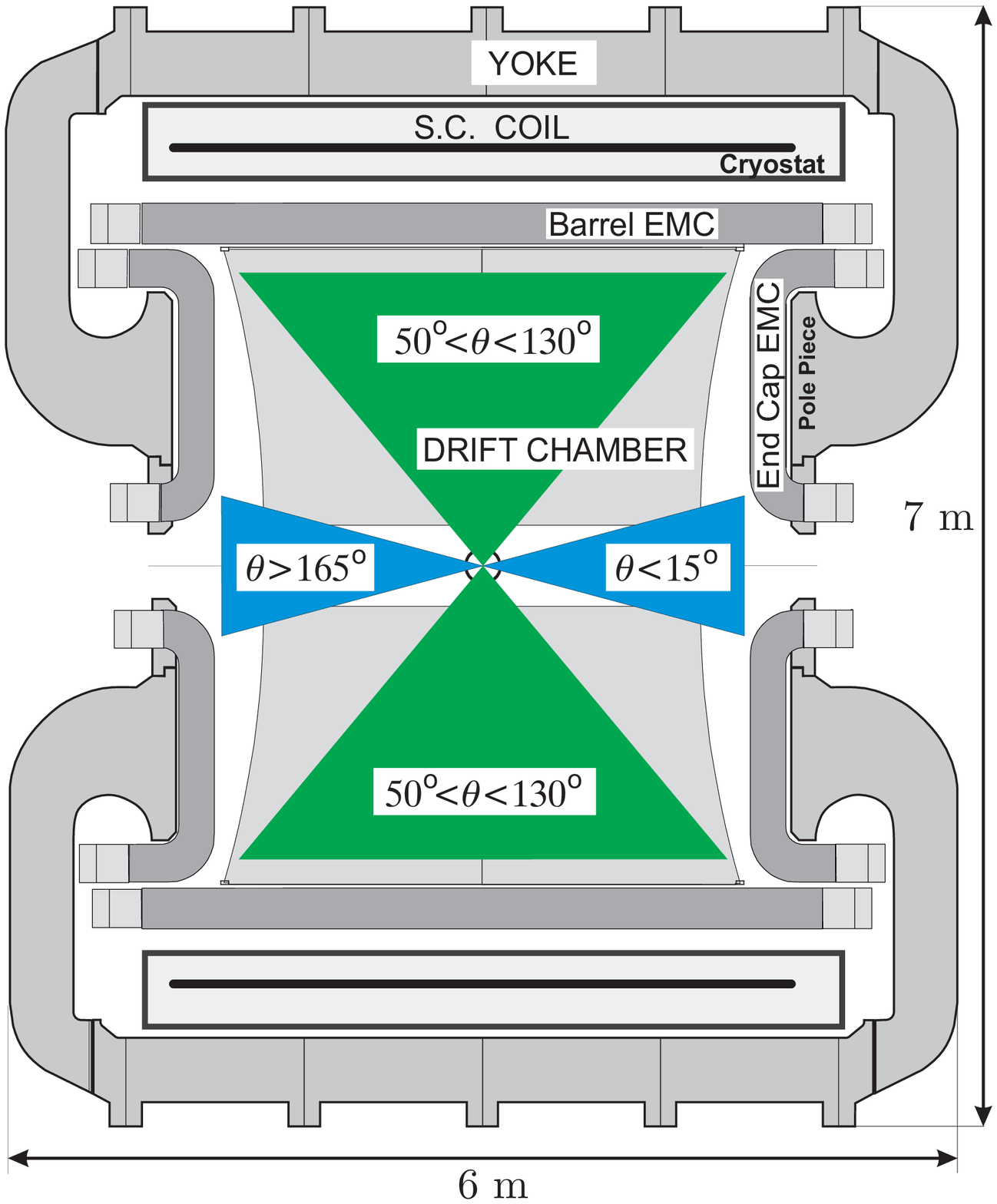}}
    \hglue 10 mm
%    \subfigure
    \resizebox{12pc}{!}{\includegraphics{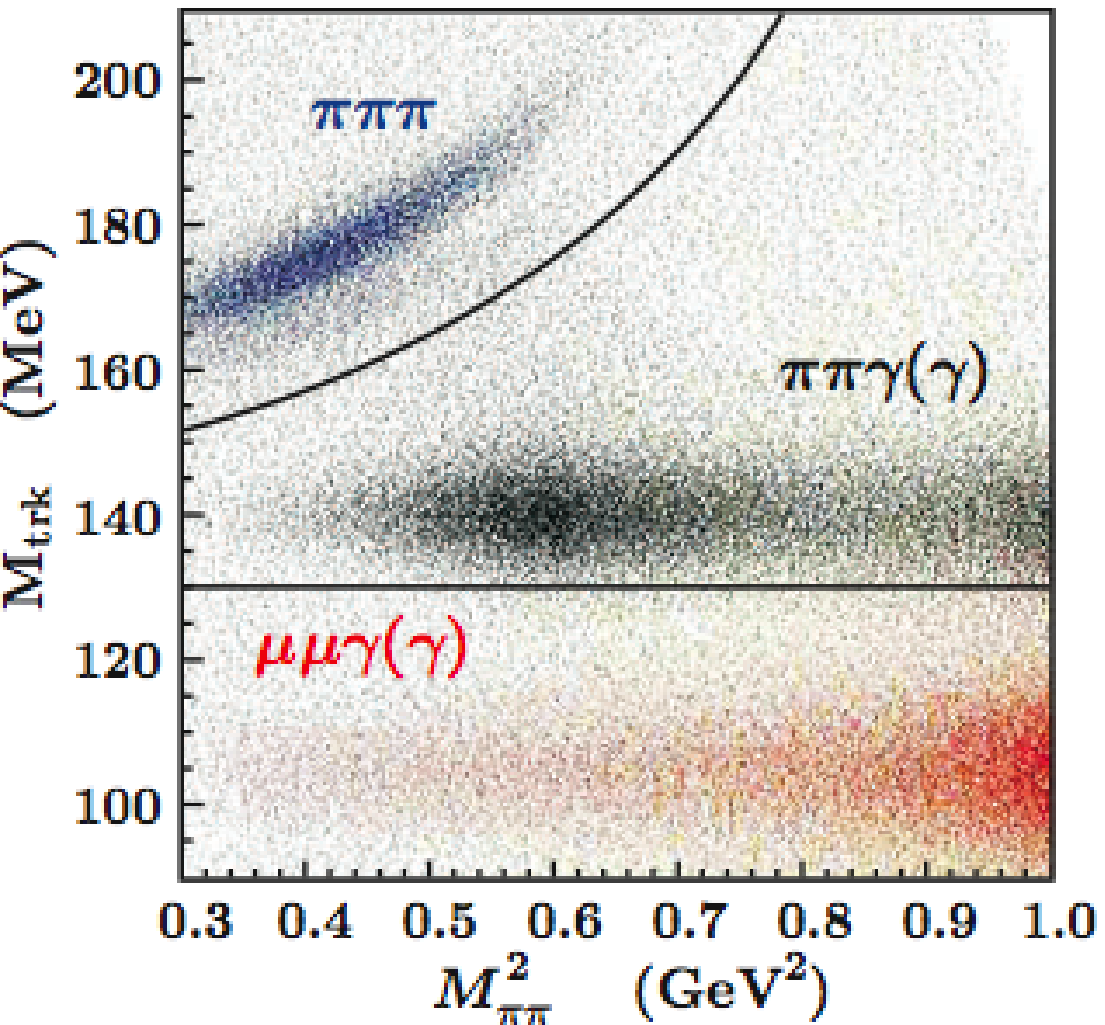}}
    \caption{\label{fig:kloe} Left: Fiducial volume for the small angle photon (narrow cones) and for the the pion tracks (wide cones). Right: Signal and background distributions in the $M_{\rm Trk}$-$M^2_{\pi\pi}$ plane; the selected area is shown.}
%  \end{center}
\end{figure}
% -----
\noindent

In the {\it small angle} analysis, photons are emitted within a cone of $\theta_\gamma<15^\circ$ around the 
beam line (narrow blue cones in Fig.~\ref{fig:kloe} left). The two charged pion tracks have $50^\circ<\theta_\pi<130^\circ$. The photon is not explicitly detected and its direction is reconstructed by closing the kinematics: $\vec{p}_\gamma\simeq\vec{p}_{miss}= -(\vec{p}_{\pi^+}+\vec{p}_{\pi^-})$. The separation of pion and photon selection regions greatly reduces the contamination from the resonant process $e^+e^-\to \phi\to\pi^+\pi^-\pi^0$, in which the $\pi^0$ mimics the missing momentum of the photon(s) and from the final state 
radiation process $e^+e^-\to \pi^+\pi^-\gamma_{\rm FSR}$. Since ISR-photons are mostly collinear with the beam line, a high statistics for the ISR signal events remains. On the other hand, a highly energetic photon emitted at small angle forces the pions also to be at small angles (and thus outside the selection cuts), resulting in a kinematical suppression of events with $M^2_{\pi\pi}< 0.35$ GeV$^2$. Residual contamination from the processes $\phi\to\pi^+\pi^-\pi^0$
and $e^+e^- \to\mu^+\mu^-\gamma$ are rejected by cuts in the kinematical variable {\it trackmass},~\footnote{Defined under the hypothesis that the final state consists of two charged particles with equal mass $M_{\rm Trk}$ and one photon.} see Fig.~\ref{fig:kloe} right. A particle ID estimator, based on calorimeter information and time-of-flight, is used to suppress the high rate of radiative Bhabhas.     

\section{Evaluation of $|F_\pi|^2$ and $a_\mu^{\pi\pi}$}
The $\pi\pi\gamma$ differential cross section is obtained from the observed spectrum,
$N_{obs}$, after subtracting
the re\-si\-dual background events, $N_{bkg}$, and correcting for
the selection efficiency, $\varepsilon_{sel}(M_{\pi\pi}^2)$,
and the luminosity, $\mathcal{L}$:
\begin{equation}
\frac{\dd\sigma_{\pi\pi\gamma}}
{\dd M_{\pi\pi}^2} = \frac{N_{obs}-N_{bkg}}
{\Delta M_{\pi\pi}^2}\, \frac{1}{\varepsilon_{sel}(M_{\pi\pi}^2)~ \mathcal{L}}~ .
\label{eq:2}
\end{equation}
%where mass resolution allows to have bins of
%width $\Delta M_{\pi\pi}^2=0.01\GeV^2$.
In order to correct for resolution effects,
the differential cross section is unfolded using the Bayesian method described in \cite{D'Agostini:1994zf}.
% Due to the excellent resolution of the KLOE DC ($\delta M_{\pi\pi}^2 = 2\times 10^{-3}$ GeV$^2$, 
%$\delta M_{\pi\pi}$ = 1.3  MeV) and to the choice of the bin width ($\Delta M_{\pi\pi}^2 = 0.01$ GeV$^2$) the unfolding procedure gives appreciable effects only in the $\rho-\omega$ region, $M_{\pi\pi}^2 = 6$ GeV$^2$, and does not introduce any additional systematic error on  $a^{\pi\pi}_\mu$.
The integrated luminosity, $\mathcal{L}$, is
obtained~\cite{Ambrosino:2006te} from the observed number
of Bhabha events, divided by the
effective cross section evaluated from the Monte Carlo generator
\texttt{Babayaga@NLO}~\cite{Carloni Calame:2000pz,Balossini:2006wc}.
 %with a theoretical  uncertainty 0.1\%. The experimental fractional
%uncertainty on $\mathcal{L}$ is 0.3\%.

%Since the quantity $M_{\pi\pi}$ is computed from measured momenta of the pions, it is shifted by radiative
%effects from the mass value at the $\pi^+\pi^-\gamma$ vertex, $M^0_{\pi\pi}$ (virtual photon mass).
The cross section $\sigma_{\pi\pi}(M^0_{\pi\pi})$ is obtained by 
accounting for final state emission (which shifts $M_{\pi\pi}$ to the virtual photon mass $M^0_{\pi\pi}$)
and dividing the $\pi^+\pi^-\gamma$ cross section by the radiator function $H$
(obtained from \texttt{Phokhara}~\cite{Rodrigo:2001jr,Kuhn:2002xg,Rodrigo:2001kf,Czyz:2002np,Czyz:2003ue} by setting pion form factor $F_\pi=1$) as in Eq. \ref{eq:1}.

The {\it bare} cross section $\sigma^0_{\pi\pi}$, inclusive of FSR,
needed for the $a_\mu^{\pi\pi}$ dispersion integral,
is obtained after removing vacuum polarization, VP,
%running of the fine structure constant
effects~\cite{Jegerlehner:2006ju}. Tab.~\ref{tab:amu} left shows the list of fractional systematic uncertainties of $a_\mu^{\pi\pi}$ in the mass range $0.35< M^2_{\pi\pi}<0.95$ GeV$^2$.

\begin{table}[h!]
 % \begin{center}
    \begin{tabular}{cc}
      \parbox[c]{180pt}{
%        {\scriptsize
          \begin{tabular}{|l|c|}
            \hline
            Reconstruction Filter & negligible\\
            Background subtraction & 0.3 \% \\
            Trackmass/Miss. Mass & 0.2 \% \\
            $\pi$/e-ID & negligible\\
            Tracking & 0.3 \% \\
            Trigger & 0.1 \% \\
            Unfolding & negligible \\
            Acceptance ($\theta_{\rm miss}$) & 0.2 \% \\
            Acceptance ($\theta_\pi$) & negligible \\
            Software Trigger (L3) & 0.1 \% \\
            Luminosity ($0.1_{th}\oplus 0.3_{exp}$)\% & 0.3 \% \\
            $\sqrt{s}$ dependence of $H$ & 0.2 \%\\
            \hline
            Total experimental systematics & 0.6 \% \\
            \hline
            \hline
            Vacuum Polarization &  0.1 \% \\
            FSR resummation & 0.3 \% \\
            Rad. function $H$   & 0.5 \% \\
            \hline
            Total theory systematics & 0.6 \% \\
            \hline
          \end{tabular}
 %       }
      }
      &
      \parbox[c]{180pt}{
        \renewcommand{\arraystretch}{1.6}
        \setlength{\tabcolsep}{1.2mm}
 %       {\scriptsize
          \begin{tabular}{|l|c|}
            \hline
            \multicolumn{2}{|c|}{$a_\mu^{\pi\pi}\times10^{10} ~ 0.35<M_{\pi\pi}^2<0.95$ GeV$^2$ } \\
            \hline
            KLOE05~\cite{Aloisio:2004bu,Ambrosino:2007vj}  & $384.4~\pm~0.8_{\rm stat}~\pm~4.6_{\rm sys}$ \\
            \hline
            KLOE08~\cite{Ambrosino:2008en} & $387.2~\pm~0.5_{\rm stat}~\pm~3.3_{\rm sys}$ \\
            \hline
            \hline
            \multicolumn{2}{|c|}{$a_\mu^{\pi\pi}\times10^{10} ~ 0.630<M_{\pi\pi}<0.958$ GeV} \\
            \hline
            CMD-2~\cite{Akhmetshin:2006bx} & $361.5\pm 5.1$ \\
            \hline
            SND~\cite{Achasov:2006vp} & $361.0\pm 3.4$ \\
            \hline
            KLOE08~\cite{Ambrosino:2008en} & $356.7\pm 3.1$ \\
            \hline
          \end{tabular}
%        }
      }
    \end{tabular}
    \caption{Left: Systematic errors on the extraction of $a_\mu^{\pi\pi}$ in the mass range $0.35<M^2_{\pi\pi}<0.95$ GeV$^2$. Right: Comparison among $a_\mu^{\pi\pi}$ values.}
    \label{tab:amu}
%  \end{center}
\end{table}
% -----

Tab.~\ref{tab:amu} right  shows the good agreement amongst
KLOE results, and also with the published
\begin{figure}[h!]
%  \begin{center}
%    \subfigure
    \resizebox{12pc}{!}{\includegraphics{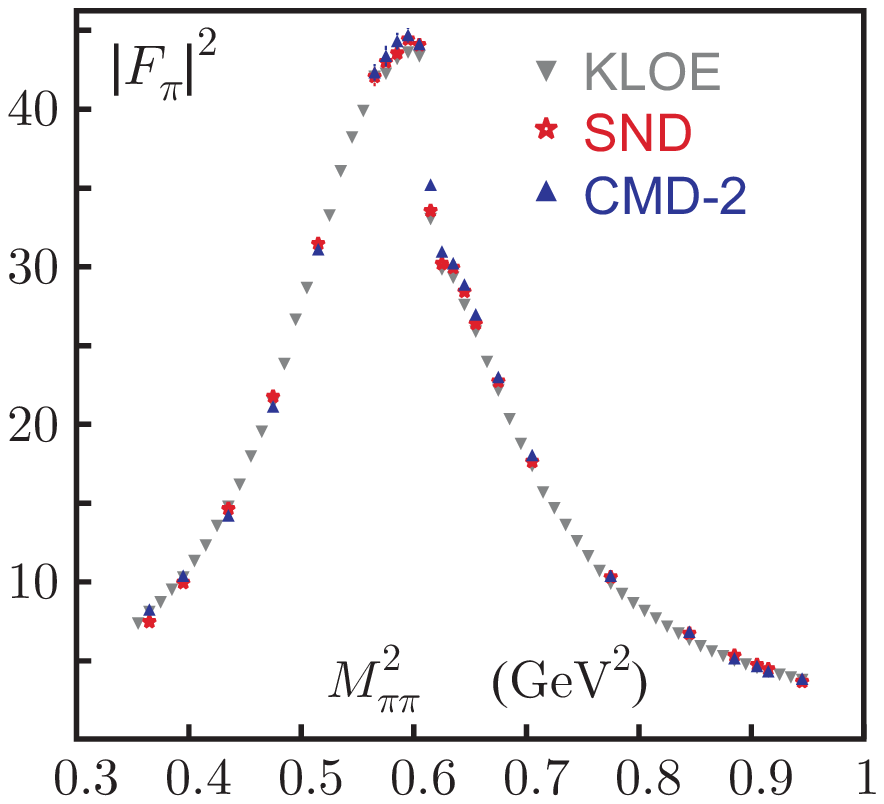}}
%    \label{fig:spp}
    \hglue 10 mm
%    \subfigure
%    \resizebox{16pc}{!}{\includegraphics[angle=90]{damu.eps}}    
    \resizebox{18pc}{!}{\includegraphics[angle=90]{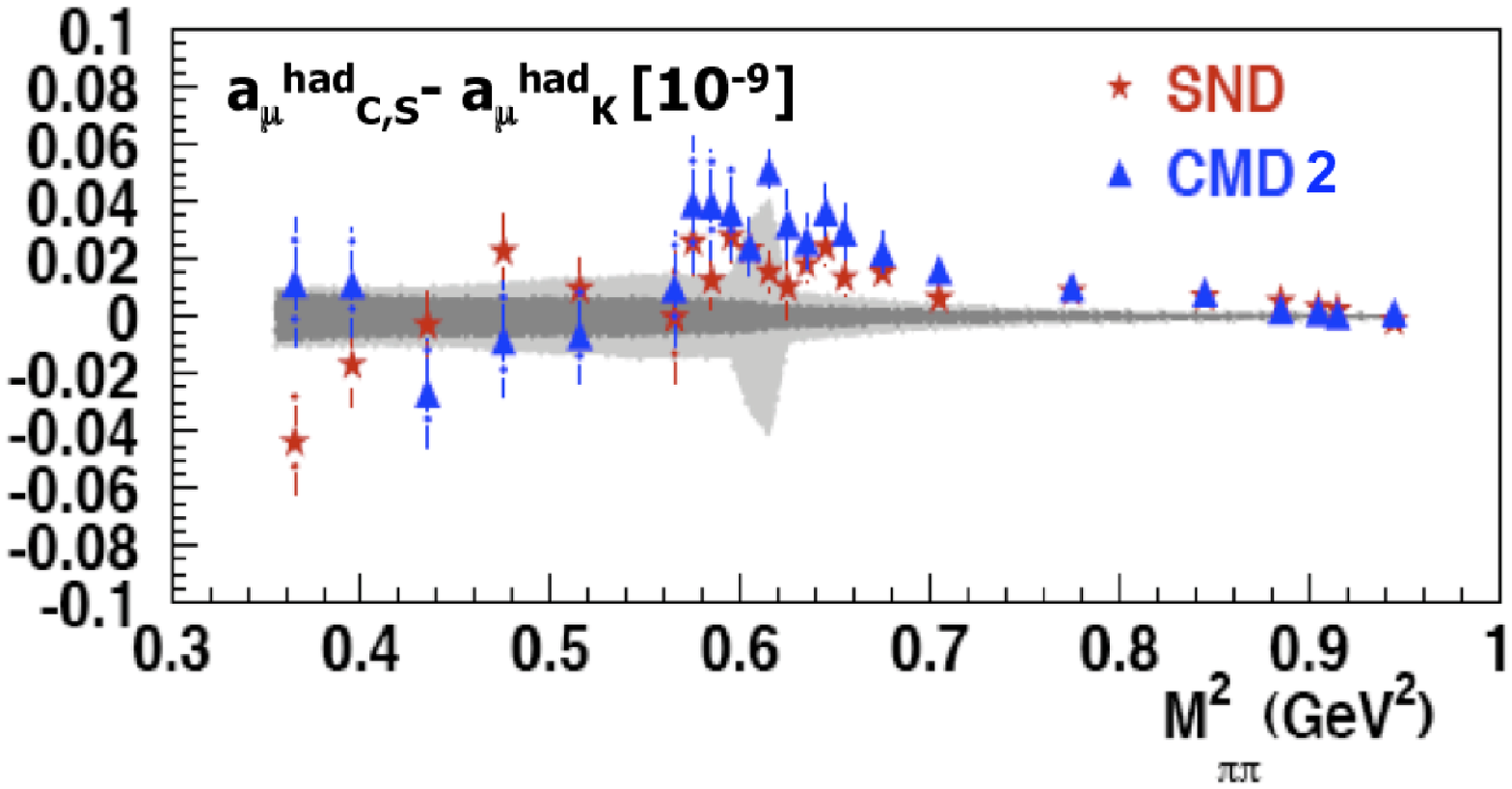}}    
    \caption{\label{fig:damu} Left: Comparison of the pion form factor measured by CMD-2, SND and
KLOE, where for this latter only statistical errors are shown. 
Right:
 Absolute difference between the dispersion integral value 
(in each energy bin) evaluated by CMD-2 or SND respect to KLOE.
The light (dark) band represents KLOE statistical (statistical$\oplus$systematic) errors.}
%  \end{center}
\end{figure}
% -----
CMD-2 and SND values. They all agree within one standard deviation.
%Due to the  improvements to the analysis described in Sec. \ref{subsec:1.3}
%and the high quality of the 2002 data, we consider the present result
%to supersede that previously published

Fig.~\ref{fig:damu} left shows a comparison of $|F_\pi|^2$ (obtained by $\sigma_{\pi\pi}$ after subtraction of FSR (assuming pointlike pions)
between CMD-2 \cite{Akhmetshin:2006bx},
  SND \cite{Achasov:2006vp} and KLOE (with only statistical errors).
For the energy scan experiments, whenever there are several data points falling in one
0.01 GeV $^2$ bin, we average the values.
Fig.~\ref{fig:damu} right shows the absolute difference the $a_\mu^{\pi\pi}$ values for each energy bin obtained in this analysis and the energy scan experiments. All the experiments are in rather good agreement within errors.

\section{Conclusions and outlook}
\begin{figure}[h]
%    \resizebox{14pc}{!}{\includegraphics{amukloe08.eps}}
    \resizebox{17pc}{!}{\includegraphics{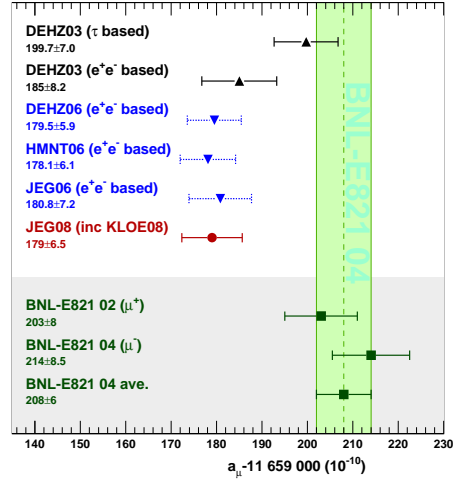}}
    \caption{\label{fig:amukloe08} Comparison of $a_{\mu}$ from theory and experiment. KLOE08 is included in JEG08~\cite{Jegerlehner:2009ry}}.
\end{figure}

KLOE has  measured the dipion contribution to the muon anomaly, $a_\mu^{\pi\pi}$, in the interval 
$0.592 < M_{\pi\pi} < 0.975$ GeV, with negligible statistical error and a 0.6\% experimental systematic uncertainty.
Theoretical uncertainties in the estimate of radiative corrections
 increase the systematic error to 0.9\%.
Combining all errors KLOE gives:
$$a_\mu^{\pi\pi}(0.592 < M_{\pi\pi} < 0.975 GeV)=(387.2\pm 3.3)\times 10^{-10}.$$
This result represents an improvement of 30\%  on the systematic error with respect to the previous published value from KLOE.
% and supersees it.
%is consistent with our previous value, with a total error smaller by 30\%.
The new result confirms the current disagreement between the standard model prediction for $a_\mu$ and
the measured value, as shown in Fig.~\ref{fig:amukloe08}.

Independent analyses are in progress to:
\begin{itemize}
\item extract the pion form factor  from
  data taken at $\sqrt{s}=1$ GeV, off the $\phi$ resonance,
  where $\pi^+\pi^-\pi^0$ background is negligible, 
  by using detected photons emitted  at large angle. This analysis, 
  which is very close to be finalized, 
  allows to measure $\sigma_{\pi\pi}$ down to the 2-pion threshold;

\item measure the pion form factor directly from the ratio, bin-by-bin,
  of $\pi^+\pi^-\gamma$ to $\mu^+\mu^-\gamma$ spectra~\cite{Muller:2006bk};

\item measure $\sigma_{\pi\pi(\gamma)}$ using the {\it large angle}
analysis at the $\phi$ peak, which would improve the knowledge of the FSR
  interference effects (in particular the $f_0(980)$ contribution~\cite{Ambrosino:2005wk,Ambrosino:2006hb}).
\end{itemize}

%\section*{Acknowledgements}
%I would like to thank my KLOE colleagues, especially P. Beltrame, C. Bini, A. Denig, W. Kluge, J.Lee Franzini, M. Moulson, S. M\"ueller, and F. Nguyen.

%Some url test \url{http://www.world.universe}.

%%%%%%%%%%%%%%%%%%%%%%%%%%%%%%%%%%%%%%%%%%%%
%% Sample figure:
%%
%% The option [height=...] scales the picture to the given height,
%% without it it would be printed at its nominal size
%%%%%%%%%%%%%%%%%%%%%%%%%%%%%%%%%%%%%%%%%%%%

%\begin{figure}
%  \includegraphics[height=.3\textheight]{golfer}
%  \caption{Picture to fixed height}
%\end{figure}

%%%%%%%%%%%%%%%%%%%%%%%%%%%%%%%%%%%%%%%%%%%%%%%%
%% BACKMATTER
%%%%%%%%%%%%%%%%%%%%%%%%%%%%%%%%%%%%%%%%%%%%%%%%

%\begin{theacknowledgments}
%  Infandum, regina, iubes renovare dolorem, Troianas ut opes et
%  lamentabile regnum cruerint Danai; quaeque ipse miserrima vidi, et
%  quorum pars magna fui. Quis talia fando Myrmidonum Dolopumve aut duri
%  miles Ulixi temperet a lacrimis?
%\end{theacknowledgments}

%\end{document}
\bibliographystyle{aipproc}   % if natbib is available
%\bibliographystyle{aipprocl} % if natbib is missing

%\endinput
%%
%% End of file `template-6s.tex'.
\end{document}